\documentclass[usenatbib, twocolumn, a4paper,nofootinbib]{aastex631}
\usepackage[utf8]{inputenc}  
\usepackage[T1]{fontenc}
\usepackage{graphicx,psfrag}
\usepackage{mathrsfs}
\usepackage{amsmath,amsfonts,amssymb,pifont,gensymb}
\usepackage{multirow,enumerate}
\usepackage{xcolor}
\usepackage{acronym}
\usepackage{xspace}
\usepackage[normalem]{ulem}
\usepackage{mathtools}
\usepackage{makecell}
\makeatletter
\let\tablewidth\relax 
\makeatother
\graphicspath{{figures/}}

\newcommand{\I}{I}

\newcommand{\N}{N}

\begin{document}
\title{A First Investigation of Repeated-Signal Localization of Strongly Lensed Gravitational Waves for Multimessenger Astronomy} 

\author[0000-0001-6728-6523]{Alvin K.Y. Li}
\email{alvinli@g.ecc.u-tokyo.ac.jp}
\affiliation{RESCEU, The University of Tokyo, Tokyo, 113-0033, Japan}

\author[0000-0002-3887-7137]{Otto A. Hannuksela}
\email{otto.akseli.hannuksela@gmail.com}
\affiliation{Department of Physics, The Chinese University of Hong Kong, Shatin, New Territories, Hong Kong}

\begin{abstract}
\noindent
Accurate sky localization is essential for gravitational-wave (GW) astronomy, particularly for multimessenger follow-up and host galaxy identification. For strongly lensed GW events, achieving localization at the level of $\sim 10~\mathrm{deg}^2$ is critical for associating signals with their lensing structures and enabling targeted searches for additional faint images. We investigate how sky localization improves when combining multiple lensed images of the same source. Using simulated lensed compact binary coalescences and \textsc{BAYESTAR} sky localization, we evaluate localization performance as a function of image multiplicity. We find that combining multiple images leads to a systematic improvement in localization, with the largest gain occurring when combining two images, typically reducing the 90\% credible region area by an order of magnitude. Additional images provide further improvements, with four-image systems achieving localization areas of $\sim 10$--$100~\mathrm{deg}^2$. We also show that subthreshold images contribute modest but non-degrading improvements, enabling their safe inclusion in localization analyses. These results demonstrate that strongly lensed GW events provide a natural pathway to improved localization and motivate hierarchical search strategies for detecting faint lensed images.

\end{abstract}

\section{Introduction}

Gravitational-wave (GW) astronomy has emerged as a powerful probe of the Universe following the first detection of a binary black hole merger by the \textit{LIGO Scientific Collaboration} and the \textit{Virgo Collaboration}~\citep{LIGOScientific:2016aoc}. Subsequent observations have established a growing population of compact binary coalescences, including binary neutron stars and neutron star--black hole systems~\citep{LIGOScientific:2021qlt,LIGOScientific:2017vwq}. These detections have enabled major advances in astrophysics, cosmology, and tests of general relativity~\citep{LIGOScientific:2016lio,LIGOScientific:2019zcs}.

A central goal of GW astronomy is accurate sky localization, which is essential for identifying electromagnetic counterparts and host galaxies~\citep{Singer:2014qca,LIGOScientific:2017apx,Fairhurst:2009tc,Fairhurst:2010is,Singer:2015ema,Singer:2016eax}. In ground-based detector networks, localization is primarily achieved through triangulation using differences in arrival time, phase, and amplitude across detectors~\citep{Fairhurst:2009tc,Fairhurst:2010is,Tsutsui:2020sml}. Rapid Bayesian localization methods such as Bayestar further enable low-latency sky maps for follow-up observations~\citep{Singer:2015ema,Tsutsui:2020sml}. The addition of more detectors improves localization by increasing the number of independent baselines and reducing geometric degeneracies~\citep{Fairhurst:2009tc,Fairhurst:2010is,Schutz:2011tw,KAGRA:2013rdx}. In particular, the inclusion of KAGRA enhances the global network by providing a geographically separated baseline and complementary antenna response~\citep{Aso:2013eba,KAGRA:2020tym}.

Despite these advances, typical sky localization uncertainties remain large, often spanning tens to hundreds of square degrees for compact binary coalescences~\citep{KAGRA:2013rdx,Fairhurst:2010is,Nissanke:2013fka,Chen:2017wpg}. This limitation poses a significant challenge for multimessenger follow-up and motivates the exploration of new approaches to improve localization performance. One promising direction is the development of hierarchical search strategies, in which information from confidently detected signals is used to guide targeted searches for additional, weaker signals, thereby improving overall detection efficiency.

Strong gravitational lensing provides a qualitatively new avenue for enhancing GW localization~\citep{Takahashi:2003ix,Nakamura:1997sw,Dai:2016igl}. When a GW signal is lensed by an intervening massive object, multiple images of the same source are produced, arriving at different times and with different magnifications~\citep{Takahashi:2003ix,Nakamura:1997sw}. Each image corresponds to an independent observation of the same intrinsic event, sharing identical source parameters but modified by lensing effects such as time delays, magnifications, and phase shifts~\citep{Dai:2016igl,Haris:2018vmn}. The detection of strongly lensed GWs has been widely anticipated in the era of advanced detectors~\citep{Ng:2017yiu,Li:2018prc,Haris:2018vmn,Hannuksela:2019kle}.

Lensed GW events have unique scientific potential. They can be used to probe cosmology through time-delay measurements~\citep{Sereno:2010dr,Liao:2017ioi,Cao:2021zpf,Hannuksela:2020xor}, test gravitational-wave propagation and related propagation effects~\citep{Collett:2016dey,Fan:2016swi,Samsing:2024xlo}, and constrain the population of compact objects and lensing structures~\citep{Oguri:2018muv,Urrutia:2021qak,Barsode:2024wda,Zumalacarregui:2024ocb,Lin:2025mpx,Prabhu:2025elp,Ubach:2025dob,Osuna:2026dzj}.. In addition, multiple images provide repeated observations of the same source, offering an opportunity to improve parameter estimation and sky localization~\citep{Hannuksela:2020xor,Liu:2020par,Janquart:2021qov,Lo:2021nae,Chen:2025xeg}.

In this work, we provide a systematic study of how strong lensing improves GW sky localization. Conceptually, each lensed image contributes an independent likelihood for the source position, and combining these likelihoods can significantly reduce the localization uncertainty. This is analogous to combining multiple independent measurements, but with additional structure arising from detector response, noise realizations, and lensing-induced correlations.

This paper is structured as follows. In Sec.~\ref{sec:background}, we review the basic properties of strongly lensed GW signals and their implications for sky localization. In Sec.~\ref{sec:method}, we describe the simulation of lensed events, detector responses, and the methodology for combining multiple images. In Sec.~\ref{sec:results}, we present our main results on localization improvement as a function of image multiplicity and assess the role of subthreshold images. Finally, in Sec.~\ref{sec:conclusion}, we summarize our findings and discuss implications for future searches and analyses.


\section{Background}\label{sec:background}

Strong gravitational lensing occurs when a gravitational wave (GW) propagates through the gravitational potential of a massive foreground object, producing multiple images of the same source. These images appear in GW detectors as repeated signals with identical intrinsic parameters but different amplitudes, arrival times, and phases~\citep{Takahashi:2003ix,Nakamura:1997sw,Dai:2016igl}.

In the geometric optics limit relevant for ground-based detectors, the waveform of each lensed image $i$ can be written in the frequency domain as
\begin{align}
\tilde{h}_i(f;\theta) = \sqrt{|\mu_i|} \, \tilde{h}_{\mathrm{src}}(f;\theta)
\exp\left(2\pi i f \Delta t_i - i \pi n_i \right),
\end{align}
where $\tilde{h}_{\mathrm{src}}(f;\theta)$ is the unlensed waveform, $\mu_i$ is the magnification factor, $\Delta t_i = \frac{1+z_L}{c} \, \Delta \phi_i$ is the time delay expressed in terms of the Fermat potential difference $\Delta \phi_i$, and $n_i$ is an integer determined by the image type, corresponding to the Morse phase associated with minima, saddle points, and maxima of the time-delay surface~\citep{Takahashi:2003ix,Nakamura:1997sw}.

The magnification rescales the observed amplitude, leading to an effective luminosity distance $D_L \propto 1/\sqrt{|\mu_i|}$ and a corresponding scaling of the signal-to-noise ratio as $\rho \propto \sqrt{|\mu_i|}$. The time delays between images arise from differences in both geometric path length and gravitational (Shapiro) delay, and typically range from minutes to weeks for galaxy-scale lenses~\citep{Takahashi:2003ix,Nakamura:1997sw,Oguri:2018muv}. In addition, images can acquire phase shifts associated with their parity, commonly expressed as multiples of $\pi/2$.

Accurate sky localization is essential for fully exploiting lensed GW events~\citep{Fairhurst:2009tc,Fairhurst:2010is,Singer:2015ema,Hannuksela:2020xor,Chen:2025xeg}. Precise localization enables the identification of host galaxies and lensing structures, which is critical for confirming lensing hypotheses, performing lens modeling, and reconstructing the lensing configuration. It also plays a central role in electromagnetic (EM) follow-up observations, which can provide independent confirmation and complementary astrophysical information.

However, current GW sky localization uncertainties remain large, often spanning tens to hundreds of square degrees for compact binary coalescences~\citep{KAGRA:2013rdx,Fairhurst:2010is,Singer:2014qca,Nissanke:2013fka}. This presents a significant challenge for associating GW events with host galaxies or lens candidates, particularly in crowded regions of the sky. Consequently, even when multiple signals are consistent with being lensed images of the same source, the lack of precise localization can hinder definitive confirmation and limit detailed lensing reconstruction.

Strongly lensed GW events offer a natural opportunity to improve sky localization~\citep{Hannuksela:2020xor,Liu:2020par,Janquart:2021qov,Lo:2021nae,Chen:2025xeg}. Each image provides an independent measurement of the same source under different noise realizations and detector responses. In ground-based detector networks, localization is primarily determined by differences in arrival time, phase, and amplitude across detectors. Each lensed image therefore contributes independent constraints on the source position through these observables, and their combination reduces the overall localization uncertainty. This is analogous to combining independent measurements, but with additional structure arising from detector response and lensing-induced effects.

Motivated by this, we investigate how sky localization can be improved by coherently combining information from multiple lensed images.


\section{Method}\label{sec:method}

In this section, we describe the methodology used to simulate strongly lensed gravitational-wave (GW) events and to evaluate the improvement in sky localization obtained by combining multiple lensed images.

\subsection{Simulation of lensed GW events}

We generate populations of strongly lensed compact binary coalescences using the \textsc{LeR} framework~\citep{Phurailatpam:2024enk}. This produces a Monte Carlo sample of lensed events with multiple images, each characterized by intrinsic source parameters (e.g., component masses and spins) and lensing-induced quantities such as magnifications and time delays~\citep{Takahashi:2003ix,Nakamura:1997sw,Li:2018prc,Ng:2017yiu}.

We select events with multiple detectable images based on signal-to-noise ratio (SNR) thresholds applied at both the single-detector and network levels~\citep{KAGRA:2013rdx,Fairhurst:2009tc,Fairhurst:2010is}. Specifically, we require a minimum single-detector SNR of $\rho_{\mathrm{th}} = 4$ and a network SNR threshold of $\rho_{\mathrm{net}} = 8$. These criteria ensure that the selected events correspond to realistic detections in current GW detector networks.

\subsection{Construction of lensed signals}

For each lensed event, we construct individual images by modifying the unlensed waveform according to the lensing parameters~\citep{Takahashi:2003ix,Nakamura:1997sw}. Following Eq.~(1), each image is generated by applying a magnification, time delay, and phase shift to the source waveform~\citep{Takahashi:2003ix,Nakamura:1997sw,Dai:2016igl}. In practice, this corresponds to:
\begin{itemize}
    \item rescaling the luminosity distance as $D_L \rightarrow D_L / \sqrt{|\mu_i|}$,
    \item shifting the coalescence time by $\Delta t_i$,
    \item applying a phase shift determined by the image type.
\end{itemize}

Each image is treated as an independent GW event while retaining the same intrinsic source parameters~\citep{Haris:2018vmn,Lo:2021nae,Janquart:2021qov}.

\subsection{Simulation of detector responses}

We simulate detector-level observables for each lensed image using the Bayestar realization pipeline~\citep{Singer:2015ema}\footnote{In reality, detection of compact binary coalescences is performed using low-latency matched-filter pipelines such as GstLAL and PyCBC~\citep{Messick:2016aqy,Nitz:2018rgo}.}. Specifically, we use the \texttt{bayestar-realize-coincs} tool to generate coincident event data based on the injected signals and assumed detector noise properties.

This produces, for each image, detector-dependent quantities including arrival times, SNRs, and phases consistent with the network response~\citep{Fairhurst:2009tc,Fairhurst:2010is,Singer:2015ema}. Gaussian noise realizations are included to model measurement uncertainties. Subthreshold triggers are retained to allow weak images to be incorporated in the analysis~\citep{Haris:2018vmn,Lo:2021nae,Janquart:2021qov}.

The resulting coincidence files (\texttt{coinc.xml}) are used as input for sky localization.

\subsection{Sky localization}

We perform sky localization for each image using the Bayestar algorithm~\citep{Singer:2015ema}, a rapid Bayesian localization method based on timing, phase, and amplitude consistency across the detector network~\citep{Fairhurst:2009tc,Fairhurst:2010is,Veitch:2014wba}. Sky maps are generated using \texttt{bayestar-localize-coincs}, which computes the posterior probability distribution over sky position using timing, phase, and amplitude consistency across detectors~\citep{Singer:2015ema,Fairhurst:2009tc,Fairhurst:2010is,Tsutsui:2020sml}.

Localization performance is quantified using credible regions, in particular the 50\% and 90\% sky areas~\citep{Singer:2015ema,KAGRA:2013rdx}.

\subsection{Combination of multiple images}

To evaluate localization improvement, we combine sky maps from multiple lensed images. Since each image provides an independent measurement of the source position under different noise realizations~\citep{Lo:2021nae,Janquart:2021qov,Chen:2025xeg,Thrane:2019xws}, the combined posterior is approximated as
\begin{align}
p_{\mathrm{comb}}(\Omega) \propto \prod_i p_i(\Omega),
\end{align}
where $p_i(\Omega)$ is the posterior for image $i$ and $\Omega$ denotes sky position.

The combined posterior is normalized to unity. This procedure aggregates information from multiple observations and reduces localization uncertainty.

\subsection{Evaluation of localization improvement}

We quantify localization improvement by comparing the 90\% credible region areas obtained from individual images and from combinations of multiple images. Localization performance is assessed by examining how $A_{90}$ changes with the number of images and by analyzing its distribution across the simulated population.

\subsubsection{Detection criteria and SNR definitions}

For each image, we compute the detector SNRs $\rho_I$ and the network SNR
\begin{align}
\rho_{\mathrm{net}} = \left( \sum_I \rho_I^2 \right)^{1/2}.
\end{align}

We classify images as follows:
\begin{itemize}
    \item \textbf{Detectable image:} $\rho_{\mathrm{net}} \geq \rho_{\mathrm{net}}^{\mathrm{th}} = 8$.
    \item \textbf{Superthreshold image:} at least one detector with $\rho_I \geq 4$, and the corresponding network SNR (computed using those detectors) satisfies $\rho_{\mathrm{net}} \geq 8$.
    \item \textbf{Subthreshold image:} images that do not meet the above criteria but are retained if they contribute to a coincident event or combination.
\end{itemize}

This classification distinguishes confidently detected images from weaker signals that may still contribute useful localization information.

\subsubsection{Single-image localization}

For each lensed event, we compute sky localization for individual images. We define the \textit{best single-image localization} as the image that yields the smallest $A_{90}$ among all images of the same source.

\subsubsection{Multi-image combinations}

We construct combined sky maps by multiplying posteriors from selected subsets of images. We consider:
\begin{itemize}
    \item \textbf{Superthreshold combinations:} using only superthreshold images,
    \item \textbf{Inclusive combinations:} including both superthreshold and subthreshold images.
\end{itemize}

For a given number of images $N$, we evaluate all available subsets of size $N$ and define the \textit{best $N$-image localization} as the combination yielding the smallest $A_{90}$. This corresponds to an optimistic estimate of achievable localization performance.


\section{Results}\label{sec:results}

\subsection{Overall improvement from multiple images}

We quantify how sky localization improves as multiple lensed images are combined, using the 90\% credible region area ($A_{90}$) as our primary metric. We refer to combinations that include both superthreshold and subthreshold images as \textit{inclusive} combinations. Across all configurations, $A_{90}$ decreases monotonically with increasing number of images, demonstrating that each additional image provides independent localization information.

We also assess performance relative to the $\sim 10~\mathrm{deg}^2$ threshold~\citep{Abbott:2020abc753}, relevant for host galaxy identification and multimessenger follow-up. Combining multiple images substantially increases the fraction of events reaching this regime.

\subsection{Two-image systems: baseline improvement}

\begin{figure}[h!]
    \centering
    \includegraphics[width=0.48\textwidth]{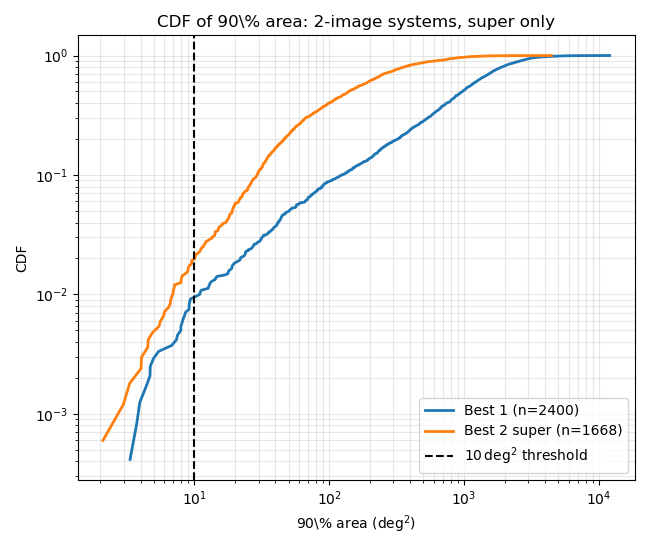}
    \includegraphics[width=0.48\textwidth]{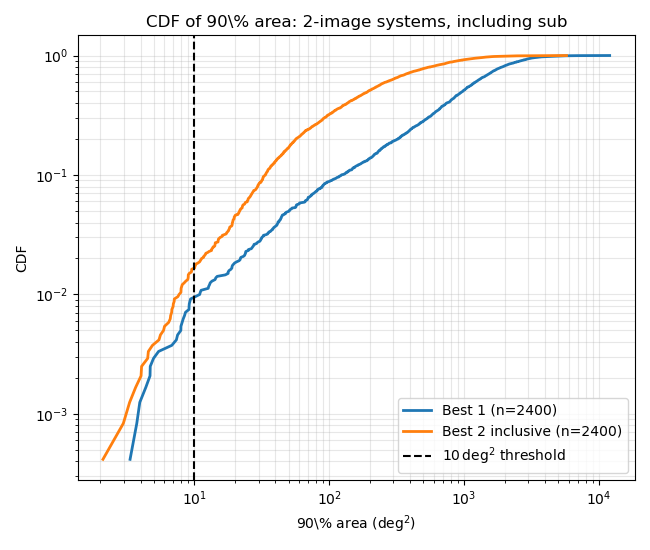}
    \caption{CDF of the 90\% credible region area for two-image systems. \textit{Left:} superthreshold-only combinations. \textit{Right:} inclusive combinations allowing subthreshold images. The dashed vertical line indicates the $10~\mathrm{deg}^2$ threshold.}
    \label{fig:2img}
\end{figure}

The results for systems with two images are shown in Fig.~\ref{fig:2img}. Combining two images leads to a clear improvement over the best single-image localization, with the distribution of $A_{90}$ shifting toward smaller values.

In the inclusive case, both curves are evaluated on the same set of systems, allowing a direct comparison. The addition of a second image produces a consistent improvement across the population, demonstrating that even a single additional image provides significant localization gain.

In the superthreshold-only case, both images must be detectable, introducing a selection bias toward higher-SNR systems. The observed improvement therefore reflects both the combination of independent measurements and the preferential inclusion of intrinsically better-localized events.

Including subthreshold images leads to similar or modestly improved localization, with no evidence of degradation. This indicates that even weak images provide additional, coherent constraints.

\subsection{Scaling with number of images}

\begin{figure}[h!]
    \centering
    \includegraphics[width=0.48\textwidth]{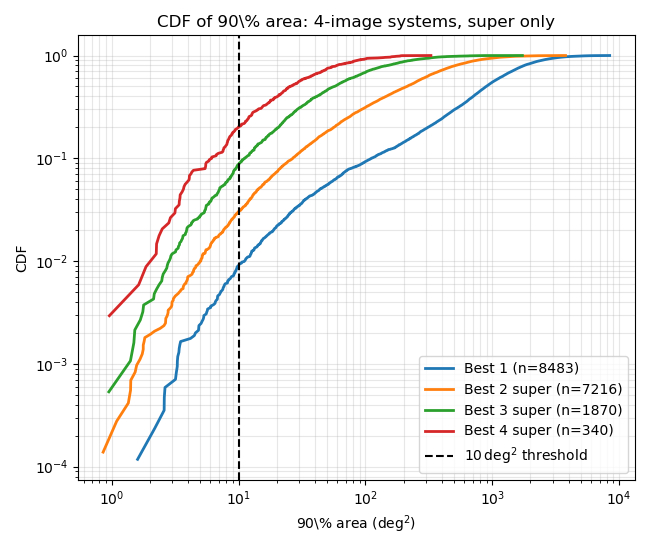}
    \includegraphics[width=0.48\textwidth]{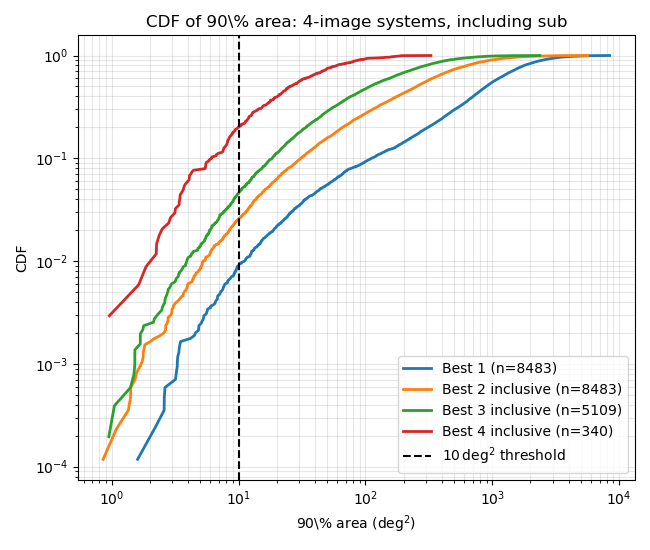}
    \caption{CDF of the 90\% credible region area for four-image systems.}
    \label{fig:4img}
\end{figure}

\begin{figure}[h!]
    \centering
    \includegraphics[width=0.48\textwidth]{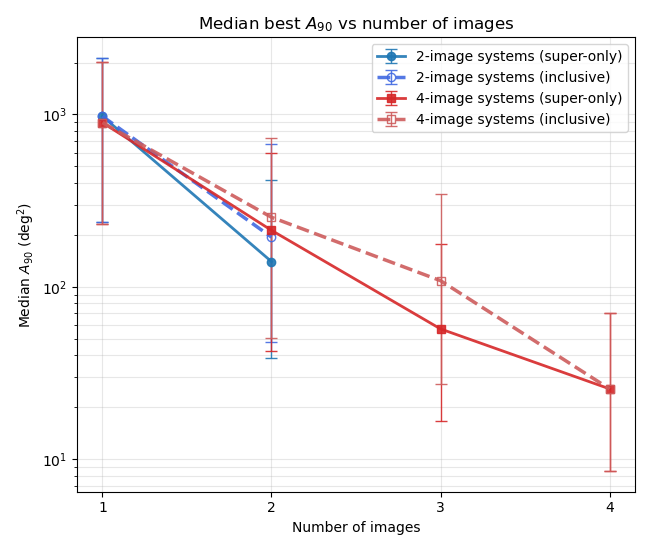}
    \caption{Median localization area as a function of the number of images. Blue curves correspond to two-image systems and red curves to four-image systems. Solid lines show superthreshold-only results, and dashed lines include subthreshold images. Error bars indicate the 16th--84th percentile range.}
    \label{fig:median_vs_n}
\end{figure}

We now examine how localization performance scales with the number of combined images, focusing on systems with up to four images. The results are shown in Fig.~\ref{fig:4img} and summarized in Fig.~\ref{fig:median_vs_n}.

Localization improves monotonically as the number of images increases from one to four. The largest gain occurs when combining two images, corresponding to an order-of-magnitude reduction in typical sky area. This reflects the transition from a single observation to multiple independent constraints on the source position.

Additional improvements from the third and fourth images are smaller but remain significant, indicating that localization does not saturate at two images. Each additional image contributes independent geometric constraints that further refine the sky position.

As shown in Fig.~\ref{fig:median_vs_n}, the median $A_{90}$ decreases steadily with image multiplicity, reaching $\sim 10$--$100~\mathrm{deg}^2$ for four-image systems. This corresponds to nearly two orders of magnitude improvement relative to single-image localization. Including subthreshold images produces a similar trend, with slight improvements in some regimes but no change in overall behavior.

\subsection{Distribution of localization performance}

\begin{figure}[h!]
    \centering
    \includegraphics[width=0.48\textwidth]{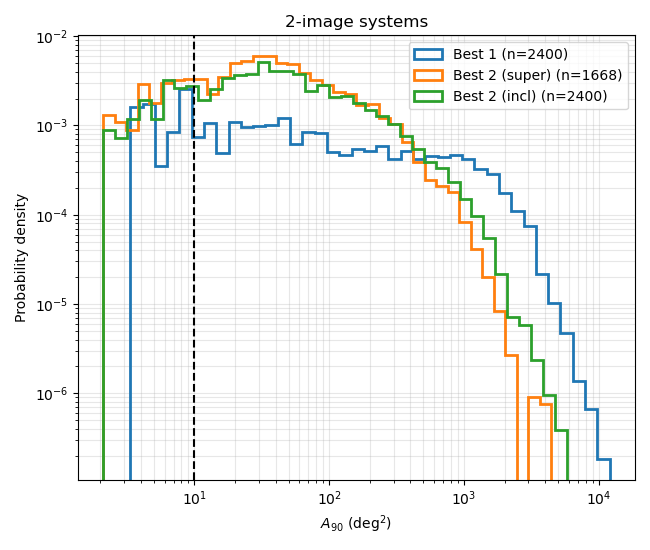}
    \caption{Distribution of $A_{90}$ for two-image systems.}
    \label{fig:hist_2img}
\end{figure}

\begin{figure}[h!]
    \centering
    \includegraphics[width=0.48\textwidth]{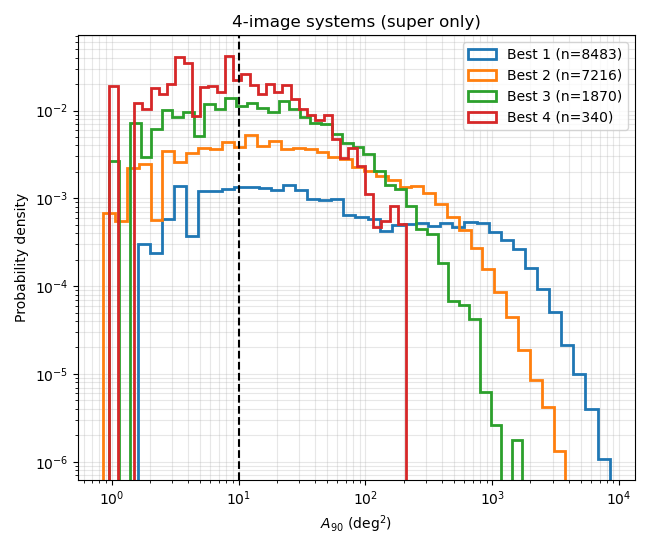}
    \includegraphics[width=0.48\textwidth]{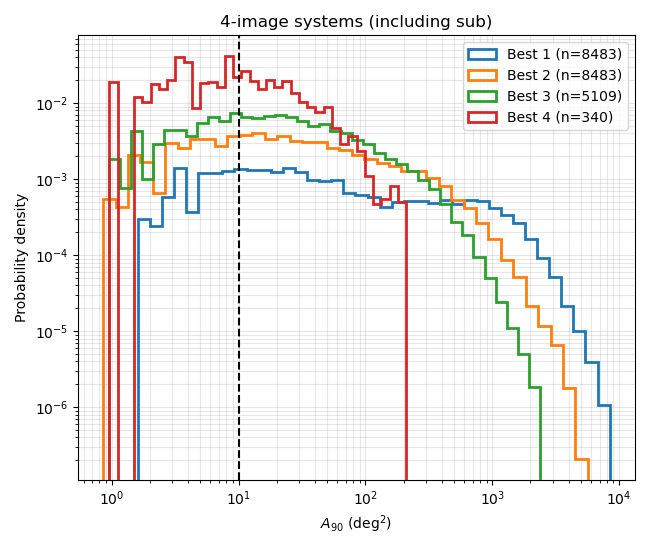}
    \caption{Distribution of $A_{90}$ for four-image systems.}
    \label{fig:hist_4img}
\end{figure}

We further examine the full distribution of $A_{90}$, as shown in Figs.~\ref{fig:hist_2img} and~\ref{fig:hist_4img}.

For two-image systems, the distribution remains broad, with a peak around $\sim 10^2$--$10^3~\mathrm{deg}^2$ and a long tail toward larger areas. While combining two images improves the typical localization, significant event-to-event variability remains.

For higher-multiplicity systems, the distribution shifts toward smaller areas and becomes more concentrated. Four-image combinations cluster around $\sim 10$--$100~\mathrm{deg}^2$, indicating both improved localization and reduced variance. Thus, combining multiple images improves not only the typical localization area but also its consistency across events.

Inclusive combinations closely follow the superthreshold-only distributions, with slight shifts toward smaller areas in some cases. This confirms that subthreshold images provide additional but subdominant information, improving localization without degradation.

\subsection{Summary of key findings}

In summary:
\begin{enumerate}
    \item Combining two images yields the largest improvement, reducing localization areas by roughly an order of magnitude.
    \item Additional images provide continued, though diminishing, improvements, with no evidence of saturation up to four images.
    \item Four-image systems achieve localization areas of $\sim 10$--$100~\mathrm{deg}^2$, approaching the regime required for host identification.
    \item Subthreshold images provide modest but consistently non-degrading improvements and can be safely included.
\end{enumerate}


\section{Conclusion}\label{sec:conclusion}

In this work, we have investigated how strong gravitational lensing can improve gravitational-wave (GW) sky localization by combining information from multiple lensed images. Using simulated lensed events and Bayestar sky localization, we quantified localization performance as a function of image multiplicity and signal detectability.

We find that combining multiple lensed images leads to a systematic and monotonic improvement in sky localization. The largest gain occurs when combining two images, which typically reduces the 90\% credible region area by an order of magnitude compared to single-image localization. Additional images provide continued, though diminishing, improvements, with no evidence of saturation up to four images. In particular, four-image systems routinely achieve localization areas of $\sim 10$--$100~\mathrm{deg}^2$, approaching the regime required for host galaxy identification and efficient electromagnetic follow-up.

We also find that subthreshold images contribute modest but consistently non-degrading improvements. Although their individual signal-to-noise ratios are low, they provide additional coherent constraints that refine the combined posterior. This demonstrates that weak images can be safely incorporated into localization analyses without introducing systematic degradation.

These results have important implications for targeted subthreshold searches, such as TESLA-X~\citep{Li:2023zdl,Li:2019osa} and pyCBC sub-threshold search~\citep{McIsaac:2019use}. 
In particular, our results provide direct quantitative support for such frameworks: improved sky localization from confidently detected images reduces the effective parameter space for subsequent searches, thereby lowering trials factors and background contamination. 
This enables a significant gain in sensitivity to subthreshold signals, which would otherwise remain undetectable in all-sky searches. This suggests a natural and potentially optimal hierarchical search strategy in which well-localized events guide targeted searches for additional lensed images, improving both detection efficiency and statistical significance.

These results highlight the unique advantage of strongly lensed GW events as natural multi-observation systems. By effectively combining independent measurements of the same source, lensing enables significant improvements in localization beyond what is achievable with single detections alone.

Looking forward, several extensions of this work are already in progress. First, we are developing a joint localization framework that extends Bayestar to coherently combine multiple events while incorporating additional physical constraints, such as relative phase shifts (e.g., Morse phase), magnification ratios, and time delays informed by lens models. Second, we are investigating coherence-based statistics for identifying lensed events by comparing the consistency of multiple signals. In particular, the distribution of coherence ratios for true lensed images (foreground) and unrelated events or noise (background) may provide a powerful discriminator for filtering candidates in targeted subthreshold searches and guiding follow-up analyses.

As GW detector sensitivity improves and the detection of lensed events becomes more likely, multi-image localization and targeted follow-up strategies are expected to play an increasingly important role in enabling precise source identification, multimessenger astronomy, and lensing-based cosmological studies. More broadly, this work suggests that strongly lensed gravitational-wave events should not be treated as independent detections, but rather as components of a structured multi-observation system, enabling new analysis strategies that go beyond conventional single-event inference.

\section*{Acknowledgements}
K.Y.A.L. would like to acknowledge the support from the Croucher Foundation.
O.A.H. acknowledge support by grants from the Research Grants Council of Hong Kong (Project No. CUHK 14304622, 14307923, and 14307724), the start-up grant from the Chinese University of Hong Kong, and the Direct Grant for Research from the Research Committee of The Chinese University of Hong Kong. 

This research has made use of data or software obtained from the Gravitational Wave Open Science Center (gwosc.org)~\citep{LIGOScientific:2025snk,KAGRA:2023pio,LIGOScientific:2019lzm}, a service of the LIGO Scientific Collaboration, the Virgo Collaboration, and KAGRA. This material is based upon work supported by NSF's LIGO Laboratory which is a major facility fully funded by the National Science Foundation, as well as the Science and Technology Facilities Council (STFC) of the United Kingdom, the Max-Planck-Society (MPS), and the State of Niedersachsen/Germany for support of the construction of Advanced LIGO and construction and operation of the GEO600 detector. Additional support for Advanced LIGO was provided by the Australian Research Council. Virgo is funded, through the European Gravitational Observatory (EGO), by the French Centre National de Recherche Scientifique (CNRS), the Italian Istituto Nazionale di Fisica Nucleare (INFN) and the Dutch Nikhef, with contributions by institutions from Belgium, Germany, Greece, Hungary, Ireland, Japan, Monaco, Poland, Portugal, Spain. KAGRA is supported by Ministry of Education, Culture, Sports, Science and Technology (MEXT), Japan Society for the Promotion of Science (JSPS) in Japan; National Research Foundation (NRF) and Ministry of Science and ICT (MSIT) in Korea; Academia Sinica (AS) and National Science and Technology Council (NSTC) in Taiwan. 

\bibliography{bibliography}

@article{LIGOScientific:2016aoc,
    author = "Abbott, B. P. and others",
    collaboration = "LIGO Scientific, Virgo",
    title = "{Observation of Gravitational Waves from a Binary Black Hole Merger}",
    eprint = "1602.03837",
    archivePrefix = "arXiv",
    primaryClass = "gr-qc",
    reportNumber = "LIGO-P150914",
    doi = "10.1103/PhysRevLett.116.061102",
    journal = "Phys. Rev. Lett.",
    volume = "116",
    number = "6",
    pages = "061102",
    year = "2016"
}

@article{LIGOScientific:2021qlt,
    author = "Abbott, R. and others",
    collaboration = "LIGO Scientific, KAGRA, VIRGO",
    title = "{Observation of Gravitational Waves from Two Neutron Star{\textendash}Black Hole Coalescences}",
    eprint = "2106.15163",
    archivePrefix = "arXiv",
    primaryClass = "astro-ph.HE",
    reportNumber = "LIGO Document P2000357",
    doi = "10.3847/2041-8213/ac082e",
    journal = "Astrophys. J. Lett.",
    volume = "915",
    number = "1",
    pages = "L5",
    year = "2021"
}

@article{LIGOScientific:2017vwq,
    author = "Abbott, B. P. and others",
    collaboration = "LIGO Scientific, Virgo",
    title = "{GW170817: Observation of Gravitational Waves from a Binary Neutron Star Inspiral}",
    eprint = "1710.05832",
    archivePrefix = "arXiv",
    primaryClass = "gr-qc",
    reportNumber = "LIGO-P170817",
    doi = "10.1103/PhysRevLett.119.161101",
    journal = "Phys. Rev. Lett.",
    volume = "119",
    number = "16",
    pages = "161101",
    year = "2017"
}

@article{LIGOScientific:2016lio,
    author = "Abbott, B. P. and others",
    collaboration = "LIGO Scientific, Virgo",
    title = "{Tests of general relativity with GW150914}",
    eprint = "1602.03841",
    archivePrefix = "arXiv",
    primaryClass = "gr-qc",
    reportNumber = "LIGO-P1500213",
    doi = "10.1103/PhysRevLett.116.221101",
    journal = "Phys. Rev. Lett.",
    volume = "116",
    number = "22",
    pages = "221101",
    year = "2016",
    note = "[Erratum: Phys.Rev.Lett. 121, 129902 (2018)]"
}

@article{LIGOScientific:2019zcs,
    author = "Abbott, B. P. and others",
    collaboration = "LIGO Scientific, Virgo, VIRGO",
    title = "{A Gravitational-wave Measurement of the Hubble Constant Following the Second Observing Run of Advanced LIGO and Virgo}",
    eprint = "1908.06060",
    archivePrefix = "arXiv",
    primaryClass = "astro-ph.CO",
    reportNumber = "LIGO-P1900015",
    doi = "10.3847/1538-4357/abdcb7",
    journal = "Astrophys. J.",
    volume = "909",
    number = "2",
    pages = "218",
    year = "2021"
}

@article{Singer:2014qca,
    author = "Singer, Leo P. and others",
    title = "{The First Two Years of Electromagnetic Follow-Up with Advanced LIGO and Virgo}",
    eprint = "1404.5623",
    archivePrefix = "arXiv",
    primaryClass = "astro-ph.HE",
    reportNumber = "LIGO-P1300187-V22, LIGO-P1300187-V24, LIGO-P1300187-V25",
    doi = "10.1088/0004-637X/795/2/105",
    journal = "Astrophys. J.",
    volume = "795",
    number = "2",
    pages = "105",
    year = "2014"
}

@article{LIGOScientific:2017apx,
    author = "Abbott, B. P. and others",
    collaboration = "LIGO Scientific, Virgo",
    title = "{On the Progenitor of Binary Neutron Star Merger GW170817}",
    eprint = "1710.05838",
    archivePrefix = "arXiv",
    primaryClass = "astro-ph.HE",
    reportNumber = "LIGO-P1700264",
    doi = "10.3847/2041-8213/aa93fc",
    journal = "Astrophys. J. Lett.",
    volume = "850",
    number = "2",
    pages = "L40",
    year = "2017"
}

@article{Fairhurst:2009tc,
    author = "Fairhurst, Stephen",
    title = "{Triangulation of gravitational wave sources with a network of detectors}",
    eprint = "0908.2356",
    archivePrefix = "arXiv",
    primaryClass = "gr-qc",
    doi = "10.1088/1367-2630/11/12/123006",
    journal = "New J. Phys.",
    volume = "11",
    pages = "123006",
    year = "2009",
    note = "[Erratum: New J.Phys. 13, 069602 (2011)]"
}

@article{Fairhurst:2010is,
    author = "Fairhurst, Stephen",
    title = "{Source localization with an advanced gravitational wave detector network}",
    eprint = "1010.6192",
    archivePrefix = "arXiv",
    primaryClass = "gr-qc",
    doi = "10.1088/0264-9381/28/10/105021",
    journal = "Class. Quant. Grav.",
    volume = "28",
    pages = "105021",
    year = "2011"
}

@article{Tsutsui:2020sml,
    author = "Tsutsui, Takuya and Cannon, Kipp and Tsukada, Leo",
    title = "{High speed source localization in searches for gravitational waves from compact object collisions}",
    eprint = "2005.08163",
    archivePrefix = "arXiv",
    primaryClass = "astro-ph.HE",
    doi = "10.1103/PhysRevD.103.043011",
    journal = "Phys. Rev. D",
    volume = "103",
    number = "4",
    pages = "043011",
    year = "2021"
}

@article{Singer:2015ema,
    author = "Singer, Leo P. and Price, Larry R.",
    title = "{Rapid Bayesian position reconstruction for gravitational-wave transients}",
    eprint = "1508.03634",
    archivePrefix = "arXiv",
    primaryClass = "gr-qc",
    reportNumber = "LIGO-P1500009-V3, LIGO-P1500009-V4, LIGO-P1500009-V5, LIGO-P1500009-V6, LIGO-P1500009-V7, LIGO-P1500009-V8",
    doi = "10.1103/PhysRevD.93.024013",
    journal = "Phys. Rev. D",
    volume = "93",
    number = "2",
    pages = "024013",
    year = "2016"
}

@article{Schutz:2011tw,
    author = "Schutz, Bernard F.",
    title = "{Networks of gravitational wave detectors and three figures of merit}",
    eprint = "1102.5421",
    archivePrefix = "arXiv",
    primaryClass = "astro-ph.IM",
    reportNumber = "AEI-2011-008",
    doi = "10.1088/0264-9381/28/12/125023",
    journal = "Class. Quant. Grav.",
    volume = "28",
    pages = "125023",
    year = "2011"
}

@article{Aso:2013eba,
    author = "Aso, Yoichi and Michimura, Yuta and Somiya, Kentaro and Ando, Masaki and Miyakawa, Osamu and Sekiguchi, Takanori and Tatsumi, Daisuke and Yamamoto, Hiroaki",
    collaboration = "KAGRA",
    title = "{Interferometer design of the KAGRA gravitational wave detector}",
    eprint = "1306.6747",
    archivePrefix = "arXiv",
    primaryClass = "gr-qc",
    doi = "10.1103/PhysRevD.88.043007",
    journal = "Phys. Rev. D",
    volume = "88",
    number = "4",
    pages = "043007",
    year = "2013"
}

@article{KAGRA:2020tym,
    author = "Akutsu, T. and others",
    collaboration = "KAGRA",
    title = "{Overview of KAGRA: Detector design and construction history}",
    eprint = "2005.05574",
    archivePrefix = "arXiv",
    primaryClass = "physics.ins-det",
    doi = "10.1093/ptep/ptaa125",
    journal = "PTEP",
    volume = "2021",
    number = "5",
    pages = "05A101",
    year = "2021"
}

@article{KAGRA:2013rdx,
    author = "Abbott, B. P. and others",
    collaboration = "KAGRA, LIGO Scientific, Virgo",
    title = "{Prospects for observing and localizing gravitational-wave transients with Advanced LIGO, Advanced Virgo and KAGRA}",
    eprint = "1304.0670",
    archivePrefix = "arXiv",
    primaryClass = "gr-qc",
    reportNumber = "LIGO-P1200087, VIR-0288A-12, JGW-P1808427",
    doi = "10.1007/s41114-020-00026-9",
    journal = "Living Rev. Rel.",
    volume = "19",
    pages = "1",
    year = "2016"
}

@article{Takahashi:2003ix,
    author = "Takahashi, Ryuichi and Nakamura, Takashi",
    title = "{Wave effects in gravitational lensing of gravitational waves from chirping binaries}",
    eprint = "astro-ph/0305055",
    archivePrefix = "arXiv",
    doi = "10.1086/377430",
    journal = "Astrophys. J.",
    volume = "595",
    pages = "1039--1051",
    year = "2003"
}

@article{Nakamura:1997sw,
    author = "Nakamura, Takahiro T.",
    title = "{Gravitational lensing of gravitational waves from inspiraling binaries by a point mass lens}",
    reportNumber = "UTAP-272-97, YITP-97-61",
    doi = "10.1103/PhysRevLett.80.1138",
    journal = "Phys. Rev. Lett.",
    volume = "80",
    pages = "1138--1141",
    year = "1998"
}

@article{Dai:2016igl,
    author = "Dai, Liang and Venumadhav, Tejaswi and Sigurdson, Kris",
    title = "{Effect of lensing magnification on the apparent distribution of black hole mergers}",
    eprint = "1605.09398",
    archivePrefix = "arXiv",
    primaryClass = "astro-ph.CO",
    doi = "10.1103/PhysRevD.95.044011",
    journal = "Phys. Rev. D",
    volume = "95",
    number = "4",
    pages = "044011",
    year = "2017"
}

@article{Haris:2018vmn,
    author = "Haris, K. and Mehta, Ajit Kumar and Kumar, Sumit and Venumadhav, Tejaswi and Ajith, Parameswaran",
    title = "{Identifying strongly lensed gravitational wave signals from binary black hole mergers}",
    eprint = "1807.07062",
    archivePrefix = "arXiv",
    primaryClass = "gr-qc",
    reportNumber = "LIGO- P1800155",
    month = "7",
    year = "2018"
}

@article{Ng:2017yiu,
    author = "Ng, Ken K. Y. and Wong, Kaze W. K. and Broadhurst, Tom and Li, Tjonnie G. F.",
    title = "{Precise LIGO Lensing Rate Predictions for Binary Black Holes}",
    eprint = "1703.06319",
    archivePrefix = "arXiv",
    primaryClass = "astro-ph.CO",
    doi = "10.1103/PhysRevD.97.023012",
    journal = "Phys. Rev. D",
    volume = "97",
    number = "2",
    pages = "023012",
    year = "2018"
}

@article{Li:2018prc,
    author = "Li, Shun-Sheng and Mao, Shude and Zhao, Yuetong and Lu, Youjun",
    title = "{Gravitational lensing of gravitational waves: A statistical perspective}",
    eprint = "1802.05089",
    archivePrefix = "arXiv",
    primaryClass = "astro-ph.CO",
    doi = "10.1093/mnras/sty411",
    journal = "Mon. Not. Roy. Astron. Soc.",
    volume = "476",
    number = "2",
    pages = "2220--2229",
    year = "2018"
}

@article{Sereno:2010dr,
    author = "Sereno, M. and Sesana, A. and Bleuler, A. and Jetzer, Ph. and Volonteri, M. and Begelman, M. C.",
    title = "{Strong lensing of gravitational waves as seen by LISA}",
    eprint = "1011.5238",
    archivePrefix = "arXiv",
    primaryClass = "astro-ph.CO",
    doi = "10.1103/PhysRevLett.105.251101",
    journal = "Phys. Rev. Lett.",
    volume = "105",
    pages = "251101",
    year = "2010"
}

@article{Liao:2017ioi,
    author = "Liao, Kai and Fan, Xi-Long and Ding, Xu-Heng and Biesiada, Marek and Zhu, Zong-Hong",
    title = "{Precision cosmology from future lensed gravitational wave and electromagnetic signals}",
    eprint = "1703.04151",
    archivePrefix = "arXiv",
    primaryClass = "astro-ph.CO",
    doi = "10.1038/s41467-017-01152-9",
    journal = "Nature Commun.",
    volume = "8",
    number = "1",
    pages = "1148",
    year = "2017",
    note = "[Erratum: Nature Commun. 8, 2136 (2017)]"
}

@article{Cao:2021zpf,
    author = "Cao, Meng-Di and Zheng, Jie and Qi, Jing-Zhao and Zhang, Xin and Zhu, Zong-Hong",
    title = "{A New Way to Explore Cosmological Tensions Using Gravitational Waves and Strong Gravitational Lensing}",
    eprint = "2112.14564",
    archivePrefix = "arXiv",
    primaryClass = "astro-ph.CO",
    doi = "10.3847/1538-4357/ac7ce4",
    journal = "Astrophys. J.",
    volume = "934",
    number = "2",
    pages = "108",
    year = "2022"
}

@article{Hannuksela:2020xor,
    author = "Hannuksela, Otto A. and Collett, Thomas E. and {\c{C}}al{\i}{\c{s}}kan, Mesut and Li, Tjonnie G. F.",
    title = "{Localizing merging black holes with sub-arcsecond precision using gravitational-wave lensing}",
    eprint = "2004.13811",
    archivePrefix = "arXiv",
    primaryClass = "astro-ph.HE",
    doi = "10.1093/mnras/staa2577",
    journal = "Mon. Not. Roy. Astron. Soc.",
    volume = "498",
    number = "3",
    pages = "3395--3402",
    year = "2020"
}

@article{Collett:2016dey,
    author = "Collett, Thomas E. and Bacon, David",
    title = "{Testing the speed of gravitational waves over cosmological distances with strong gravitational lensing}",
    eprint = "1602.05882",
    archivePrefix = "arXiv",
    primaryClass = "astro-ph.HE",
    doi = "10.1103/PhysRevLett.118.091101",
    journal = "Phys. Rev. Lett.",
    volume = "118",
    number = "9",
    pages = "091101",
    year = "2017"
}

@article{Fan:2016swi,
    author = "Fan, Xi-Long and Liao, Kai and Biesiada, Marek and Piorkowska-Kurpas, Aleksandra and Zhu, Zong-Hong",
    title = "{Speed of Gravitational Waves from Strongly Lensed Gravitational Waves and Electromagnetic Signals}",
    eprint = "1612.04095",
    archivePrefix = "arXiv",
    primaryClass = "gr-qc",
    doi = "10.1103/PhysRevLett.118.091102",
    journal = "Phys. Rev. Lett.",
    volume = "118",
    number = "9",
    pages = "091102",
    year = "2017"
}

@article{Samsing:2024xlo,
    author = "Samsing, Johan and Zwick, Lorenz and Saini, Pankaj and D'Orazio, Daniel J. and Hendriks, Kai and Ezquiaga, Jose Mar{\'\i}a and Lo, Rico K. L. and Vujeva, Luka and Radev, Georgi D. and Yu, Yan",
    title = "{Measuring the Transverse Velocity of Strongly Lensed Gravitational Wave Sources with Ground Based Detectors}",
    eprint = "2412.14159",
    archivePrefix = "arXiv",
    primaryClass = "astro-ph.HE",
    month = "12",
    year = "2024"
}

@article{Oguri:2018muv,
    author = "Oguri, Masamune",
    title = "{Effect of gravitational lensing on the distribution of gravitational waves from distant binary black hole mergers}",
    eprint = "1807.02584",
    archivePrefix = "arXiv",
    primaryClass = "astro-ph.CO",
    doi = "10.1093/mnras/sty2145",
    journal = "Mon. Not. Roy. Astron. Soc.",
    volume = "480",
    number = "3",
    pages = "3842--3855",
    year = "2018"
}

@article{Lin:2025mpx,
    author = "Lin, Xin-Yi and Wang, Xi-Jing and Zhou, Huan and Li, Zhengxiang and Liao, Kai and Zhu, Zong-Hong",
    title = "{Constraints on Compact Dark Matter Population from Micro-lensing Effect of Gravitational Wave for the third-generation gravitational Wave Detector}",
    eprint = "2508.13577",
    archivePrefix = "arXiv",
    primaryClass = "astro-ph.CO",
    month = "8",
    year = "2025"
}

@article{Ubach:2025dob,
    author = "Ubach, Helena and Gieles, Mark and Miralda-Escud{\'e}, Jordi",
    title = "{Constraining the environment of compact binary mergers with self-lensing signatures}",
    eprint = "2505.04794",
    archivePrefix = "arXiv",
    primaryClass = "astro-ph.HE",
    doi = "10.1103/ql7q-t6wc",
    journal = "Phys. Rev. D",
    volume = "112",
    number = "8",
    pages = "083026",
    year = "2025"
}

@article{Barsode:2024wda,
    author = "Barsode, A. and Kapadia, S. J. and Ajith, P.",
    title = "{Constraints on Compact Dark Matter from the Nonobservation of Gravitational-wave Strong Lensing}",
    eprint = "2405.15878",
    archivePrefix = "arXiv",
    primaryClass = "gr-qc",
    doi = "10.3847/1538-4357/ad77c4",
    journal = "Astrophys. J.",
    volume = "975",
    number = "1",
    pages = "48",
    year = "2024"
}

@article{Urrutia:2021qak,
    author = "Urrutia, Juan and Vaskonen, Ville",
    title = "{Lensing of gravitational waves as a probe of compact dark matter}",
    eprint = "2109.03213",
    archivePrefix = "arXiv",
    primaryClass = "astro-ph.CO",
    doi = "10.1093/mnras/stab3118",
    journal = "Mon. Not. Roy. Astron. Soc.",
    volume = "509",
    number = "1",
    pages = "1358--1365",
    year = "2021"
}

@article{Prabhu:2025elp,
    author = "Prabhu, Gopalkrishna and Deka, Uddeepta and Chakraborty, Sumanta and Kapadia, Shasvath J.",
    title = "{Probing the spin of compact objects with gravitational microlensing of gravitational waves}",
    eprint = "2512.18707",
    archivePrefix = "arXiv",
    primaryClass = "gr-qc",
    month = "12",
    year = "2025"
}

@article{Zumalacarregui:2024ocb,
    author = "Zumalac{\'a}rregui, Miguel",
    title = "{Lens Stochastic Diffraction: A Signature of Compact Structures in Gravitational-Wave Data}",
    eprint = "2404.17405",
    archivePrefix = "arXiv",
    primaryClass = "gr-qc",
    month = "4",
    year = "2024"
}

@article{Osuna:2026dzj,
    author = "Osuna, Joel Cortez and Shandera, Sarah",
    title = "{From Origins to Observables: Distinguishing Dark Compact Objects with Population-Level Microlensing Signatures}",
    eprint = "2603.24862",
    archivePrefix = "arXiv",
    primaryClass = "astro-ph.CO",
    month = "3",
    year = "2026"
}

@article{Chen:2025xeg,
    author = "Chen, Zhiwei and Yu, Qingjuan and Lu, Youjun and Guo, Xiao",
    title = "{Enhanced Localization of Dark Lensed Gravitational-wave Events Enables Host Galaxy Identification and Precise Cosmological Inference}",
    eprint = "2510.12470",
    archivePrefix = "arXiv",
    primaryClass = "astro-ph.CO",
    doi = "10.3847/2041-8213/ae1226",
    journal = "Astrophys. J. Lett.",
    volume = "993",
    number = "2",
    pages = "L57",
    year = "2025"
}

@article{Liu:2020par,
    author = "Liu, Xiaoshu and Magana Hernandez, Ignacio and Creighton, Jolien",
    title = "{Identifying strong gravitational-wave lensing during the second observing run of Advanced LIGO and Advanced Virgo}",
    eprint = "2009.06539",
    archivePrefix = "arXiv",
    primaryClass = "astro-ph.HE",
    doi = "10.3847/1538-4357/abd7eb",
    journal = "Astrophys. J.",
    volume = "908",
    number = "1",
    pages = "97",
    year = "2021"
}

@article{Lo:2021nae,
    author = "Lo, Rico K. L. and Magana Hernandez, Ignacio",
    title = "{Bayesian statistical framework for identifying strongly lensed gravitational-wave signals}",
    eprint = "2104.09339",
    archivePrefix = "arXiv",
    primaryClass = "gr-qc",
    doi = "10.1103/PhysRevD.107.123015",
    journal = "Phys. Rev. D",
    volume = "107",
    number = "12",
    pages = "123015",
    year = "2023"
}

@article{Janquart:2021qov,
    author = "Janquart, Justin and Hannuksela, Otto A. and K., Haris and Van Den Broeck, Chris",
    title = "{A fast and precise methodology to search for and analyse strongly lensed gravitational-wave events}",
    eprint = "2105.04536",
    archivePrefix = "arXiv",
    primaryClass = "gr-qc",
    doi = "10.1093/mnras/stab1991",
    journal = "Mon. Not. Roy. Astron. Soc.",
    volume = "506",
    number = "4",
    pages = "5430--5438",
    year = "2021"
}

@article{Nissanke:2013fka,
    author = "Nissanke, Samaya and Holz, Daniel E. and Dalal, Neal and Hughes, Scott A. and Sievers, Jonathan L. and Hirata, Christopher M.",
    title = "{Determining the Hubble constant from gravitational wave observations of merging compact binaries}",
    eprint = "1307.2638",
    archivePrefix = "arXiv",
    primaryClass = "astro-ph.CO",
    month = "7",
    year = "2013"
}

@article{Chen:2017wpg,
    author = "Chen, Hsin-Yu and Holz, Daniel E. and Miller, John and Evans, Matthew and Vitale, Salvatore and Creighton, Jolien",
    title = "{Distance measures in gravitational-wave astrophysics and cosmology}",
    eprint = "1709.08079",
    archivePrefix = "arXiv",
    primaryClass = "astro-ph.CO",
    doi = "10.1088/1361-6382/abd594",
    journal = "Class. Quant. Grav.",
    volume = "38",
    number = "5",
    pages = "055010",
    year = "2021"
}

@article{Hannuksela:2019kle,
    author = "Hannuksela, O. A. and Haris, K. and Ng, K. K. Y. and Kumar, S. and Mehta, A. K. and Keitel, D. and Li, T. G. F. and Ajith, P.",
    title = "{Search for gravitational lensing signatures in LIGO-Virgo binary black hole events}",
    eprint = "1901.02674",
    archivePrefix = "arXiv",
    primaryClass = "gr-qc",
    reportNumber = "LIGO Document P1800297, LIGO-P1800297",
    doi = "10.3847/2041-8213/ab0c0f",
    journal = "Astrophys. J. Lett.",
    volume = "874",
    number = "1",
    pages = "L2",
    year = "2019"
}

@article{Phurailatpam:2024enk,
    author = "Phurailatpam, Hemantakumar and More, Anupreeta and Narola, Harsh and Yin, Ng Chung and Janquart, Justin and Van Den Broeck, Chris and Hannuksela, Otto Akseli and Singh, Neha and Keitel, David",
    title = "{ler : LVK (LIGO-Virgo-KAGRA collaboration) event (compact-binary mergers) rate calculator and simulator}",
    eprint = "2407.07526",
    archivePrefix = "arXiv",
    primaryClass = "astro-ph.IM",
    month = "7",
    year = "2024"
}

@article{LIGOScientific:2025snk,
    author = "Abac, A. G. and others",
    collaboration = "LIGO Scientific, VIRGO, KAGRA",
    title = "{Open Data from LIGO, Virgo, and KAGRA through the First Part of the Fourth Observing Run}",
    eprint = "2508.18079",
    archivePrefix = "arXiv",
    primaryClass = "gr-qc",
    reportNumber = "LIGO-P2500167",
    month = "8",
    year = "2025"
}

@article{KAGRA:2023pio,
    author = "Abbott, R. and others",
    collaboration = "KAGRA, VIRGO, LIGO Scientific",
    title = "{Open Data from the Third Observing Run of LIGO, Virgo, KAGRA, and GEO}",
    eprint = "2302.03676",
    archivePrefix = "arXiv",
    primaryClass = "gr-qc",
    reportNumber = "LIGO-P2200316",
    doi = "10.3847/1538-4365/acdc9f",
    journal = "Astrophys. J. Suppl.",
    volume = "267",
    number = "2",
    pages = "29",
    year = "2023"
}

@article{LIGOScientific:2019lzm,
    author = "Abbott, Rich and others",
    collaboration = "LIGO Scientific, Virgo",
    title = "{Open data from the first and second observing runs of Advanced LIGO and Advanced Virgo}",
    eprint = "1912.11716",
    archivePrefix = "arXiv",
    primaryClass = "gr-qc",
    reportNumber = "LIGO-P1900206",
    doi = "10.1016/j.softx.2021.100658",
    journal = "SoftwareX",
    volume = "13",
    pages = "100658",
    year = "2021"
}

@article{Messick:2016aqy,
  author = "Messick, Cody and others",
  title = "{Analysis framework for the prompt discovery of compact binary mergers in gravitational-wave data}",
  eprint = "1604.04324",
  archivePrefix = "arXiv",
  primaryClass = "astro-ph.IM",
  journal = "Phys. Rev. D",
  volume = "95",
  pages = "042001",
  year = "2017"
}

@article{Nitz:2018rgo,
  author = "Nitz, Alexander H. and others",
  title = "{Rapid detection of gravitational waves from compact binary mergers with PyCBC Live}",
  eprint = "1802.04370",
  archivePrefix = "arXiv",
  primaryClass = "astro-ph.IM",
  journal = "Phys. Rev. D",
  volume = "98",
  pages = "024050",
  year = "2018"
}

@article{Li:2023zdl,
    author = "Li, Alvin K. Y. and Chan, Juno C. L. and Fong, Heather and Chong, Aidan H. Y. and Weinstein, Alan J. and Ezquiaga, Jose M.",
    title = "{TESLA-X: an effective method to search for subthreshold lensed gravitational waves with a targeted population model}",
    eprint = "2311.06416",
    archivePrefix = "arXiv",
    primaryClass = "gr-qc",
    doi = "10.1093/mnras/staf1259",
    journal = "Mon. Not. Roy. Astron. Soc.",
    volume = "542",
    number = "2",
    pages = "998--1010",
    year = "2025"
}

@article{Li:2019osa,
    author = "Li, Alvin K. Y. and Lo, Rico K. L. and Sachdev, Surabhi and Chan, Juno C. L. and Lin, E. T. and Li, Tjonnie G. F. and Weinstein, Alan J.",
    collaboration = "LIGO Scientific, Virgo",
    title = "{Targeted subthreshold search for strongly lensed gravitational-wave events}",
    eprint = "1904.06020",
    archivePrefix = "arXiv",
    primaryClass = "gr-qc",
    doi = "10.1103/PhysRevD.107.123014",
    journal = "Phys. Rev. D",
    volume = "107",
    number = "12",
    pages = "123014",
    year = "2023"
}

@article{McIsaac:2019use,
    author = "McIsaac, Connor and Keitel, David and Collett, Thomas and Harry, Ian and Mozzon, Simone and Edy, Oliver and Bacon, David",
    title = "{Search for strongly lensed counterpart images of binary black hole mergers in the first two LIGO observing runs}",
    eprint = "1912.05389",
    archivePrefix = "arXiv",
    primaryClass = "gr-qc",
    reportNumber = "LIGO-P1900360",
    doi = "10.1103/PhysRevD.102.084031",
    journal = "Phys. Rev. D",
    volume = "102",
    number = "8",
    pages = "084031",
    year = "2020"
}

@article{Singer:2016eax,
    author = "Singer, Leo P. and others",
    title = "{Going the Distance: Mapping Host Galaxies of LIGO and Virgo Sources in Three Dimensions Using Local Cosmography and Targeted Follow-up}",
    eprint = "1603.07333",
    archivePrefix = "arXiv",
    primaryClass = "astro-ph.HE",
    reportNumber = "LIGO-P1500071-V4, LIGO-P1500071-V5, LIGO-P1500071-V6, LIGO-P1500071-V7",
    doi = "10.3847/2041-8205/829/1/L15",
    journal = "Astrophys. J. Lett.",
    volume = "829",
    number = "1",
    pages = "L15",
    year = "2016"
}

@article{Abbott:2020abc753,
  author = "Abbott, B. P. and others",
  title = "{Prospects for localization and identification of GW sources}",
  journal = "Astrophys. J. Lett.",
  year = "2020"
}

@article{Thrane:2019xws,
  author = "Thrane, Eric and Talbot, Colm",
  title = "{An introduction to Bayesian inference in gravitational-wave astronomy}",
  eprint = "1809.02293",
  archivePrefix = "arXiv",
  journal = "Publ. Astron. Soc. Austral.",
  year = "2019"
}

@article{Veitch:2014wba,
  author = "Veitch, John and others",
  title = "{Parameter estimation for compact binaries with ground-based detectors}",
  eprint = "1409.7215",
  archivePrefix = "arXiv",
  journal = "Phys. Rev. D",
  year = "2015"
}

\end{document}